\documentstyle{cupconf}

\ifoldfss
\else
  \ifnfssone
    \newmathalphabet{\mathit}
      \addtoversion{normal}{\mathit}{cmr}{m}{it}
      \addtoversion{bold}{\mathit}{cmr}{bx}{it}
    \newmathalphabet{\mathcal}
      \addtoversion{normal}{\mathcal}{cmsy}{m}{n}
    \else
    \ifnfsstwo
    \fi
  \fi
\fi

\def\eps{{\cal E}}
\def\fid{f_{\rm k}}
\def\gau{\gamma_1}
\def\gad{\gamma_2}
\def\mt{M_{\rm T}}
\def\parn{\par\noindent}
\def\psii{\Psi_{\rm k}}
\def\psit{\Psi_{\rm T}}
\def\qi{Q_{\rm k}}
\def\ra{r_{\rm a}}
\def\rai{r_{\rm ak}}
\def\rc{r_{\rm c}}
\def\roi{\rho _{\rm k}}
\def\roqi{\varrho _{\rm k}}
\def\hexnumber#1{\ifcase#1 0\or1\or2\or3\or4\or5\or6\or7\or8\or9\or
 A\or B\or C\or D\or E\or F\fi }

\makeatletter
\ifx\CUP@mtlplain@loaded\undefined
\else
\fi
\makeatother
 \makeatletter
 \ifx\CUP@mtlplain@loaded\undefined
   \font\tenbmi=cmmib10 at 10pt
   \font\sevenbmi=cmmib10 at 7pt
   \font\fivebmi=cmmib10 at 5pt

   \newfam\bmifam
   \textfont\bmifam=\tenbmi
   \scriptfont\bmifam=\sevenbmi
   \scriptscriptfont\bmifam=\fivebmi
   
 \fi
 \makeatother
\ifnfsstwo

\fi
\ifnfssone

\fi
\ifoldfss

\fi

\mathchardef\varLambda="0103

\makeatletter
\ifx\CUP@mtlplain@loaded\undefined
\else
\fi
\makeatother
\makeatletter
\ifx\CUP@mtlplain@loaded\undefined
  \font\tenbms=cmbsy10
  \font\sevenbms=cmbsy10 at 7pt
  \font\fivebms=cmbsy10 at 5pt
  \newfam\bmsfam
  \textfont\bmsfam=\tenbms
  \scriptfont\bmsfam=\sevenbms
  \scriptscriptfont\bmsfam=\fivebms

  \edef\bsy@{\hexnumber\bmsfam}
  \mathchardef\bnabla="0\bsy@72
\fi
\makeatother
\def\etal{\mbox{\it et al.}}

\title[Consistency of two--component galaxy models]
{Two--component galaxy models:\\phase--space constraints}

\author[L. Ciotti]%
{L\ls U\ls C\ls A\ns C\ls I\ls O\ls T\ls T\ls I$^{1,2}$}

\affiliation{$^1$Osservatorio Astronomico di Bologna, via Zamboni 33, 40126 
Bologna, ITALY\\[\affilskip]
$^2$Scuola Normale Superiore, Piazza dei Cavalieri 7, 56126 Pisa, ITALY}

\begin{document}
\ifnfssone
\else
  \ifnfsstwo
  \else
    \ifoldfss
      \let\mathcal\cal
      \let\mathrm\rm
      \let\mathsf\sf
    \fi
  \fi
\fi

\maketitle

\begin{abstract}

The properties of the analitycal phase--space distribution function (DF) of
two--component spherical self--consistent galaxy models, where one
density distribution follows the Hernquist profile, and the other a
$\gamma=0$ model, with different total masses and core radii (H0
models), presented in Ciotti (1998, C98), are here summarized.  A
variable amount of radial Osipkov--Merritt (OM) orbital anisotropy is
allowed in both components. The necessary and sufficient conditions
that the model parameters must satisfy in order to correspond to a
model where each one of the two distinct components has a positive DF
(the so--called model consistency) are analytically derived, together
with some results on the more general problem of the consistency of
two--component $\gau+\gad$ models. The possibility to add in a
consistent way a black hole (BH) at the center of radially anisotropic
$\gamma$ models is also discussed.  In the particular case of H0
models, it is proved that a globally isotropic Hernquist component is
consistent for any mass and core radius of the superimposed $\gamma=0$
halo; on the contrary, only a maximum value of the core radius is
allowed to the $\gamma=0$ component when a Hernquist halo is added.
The combined effect of halo concentration and orbital anisotropy is
successively investigated.

\end{abstract}

\firstsection 

\section{Introduction}

In the study of stellar dynamical models the fact that the Jeans
equations have a physically acceptable solution is not a sufficient
criterion for the validity of the model: the essential requirement to
be met is the positivity of the DF of each distinct component. A model
satisfying this minimal requirement is called a {\it consistent}
model.  In order to recover the DF of spherical models with
anisotropy, the OM technique has been developed (Osipkov 1979; Merritt
1985), and {\it numerically} applied (see, e.g., Ciotti \& Pellegrini
1992, CP92; Carollo, de Zeeuw, \& van der Marel 1995; Ciotti \&
Lanzoni 1997, CL97). In the OM framework, a simple approach in order
to check the consistency of spherically symmetric, multi--component
models (avoiding the recovering of the DF itself), is described in
CP92.  It is now accepted that a fraction of the mass in galaxies is
made of a dark component, whose density distribution -- albeit not
well constrained by observations -- differs from that of the visible
one (see, e.g., Bertin \etal 1994; Carollo \etal 1995; Buote \&
Canizares 1997; Gerhard \etal 1998).  Moreover, there is an increasing
evidence of the presence of massive BHs at the center of most (if not
all) elliptical galaxies (see, e.g., Harms \etal 1994; van der Marel
\etal 1997; Richstone 1998).  Unfortunately, only few examples of
two--component systems in which both the spatial density and the DF
are analytically known are at our disposition, namely the
Binney--Evans model (Binney 1991; Evans 1993), and the two--component
Hernquist model (HH model, Ciotti 1996, C96).  It is therefore of
interest the result proved in C98 that also the DF of H0 models with
OM anisotropy is completely expressible in analytical way.  This
family of models is made by the superposition of a density
distribution following the Hernquist profile (Hernquist 1990), and
another density distribution following the $\gamma=0$ profile [see
eq. (3.5)], with different total masses and core radii. OM orbital
anisotropy is allowed in both components.  Strictly related to the
last point above, is the trend shown by the numerical investigations
of CP92, i.e., the difficulty of consistently superimposing a
centrally peaked distribution to a centrally flat one.  More
specifically, CP92 showed numerically that King (King 1972) or
quasi--isothermal density profiles can not be coupled to a de
Vaucouleurs (de Vaucouleurs 1948) model, because their DFs run into
negative values near the model center. On the contrary, the DF of the
de Vaucouleurs component is qualitatively unaffected by the presence
of centrally flat halos.  From this point of view, the C96 work on HH
models is complementary to the investigation of CP92: in the HH models
the two density components are both centrally peaked, and their DF is
positive for all the possible choices of halo and galaxy masses and
concentrations (in the isotropic case).  The implications of these
findings have been not sufficiently explored. For example, one could
speculate that in presence of a centrally peaked dark matter halo,
King--like elliptical galaxies should be relatively rare, or,
viceversa, that a galaxy with a central power--law density profile
cannot have a dark halo too flat in the center. In fact observational
results on the central surface brightness profiles of elliptical
galaxies (see, e.g., Jaffe \etal 1994; M{\o}ller, Stiavelli, \&
Zeilinger 1995; Lauer \etal 1995), and bulges of spirals (Carollo \&
Stiavelli 1998), as well as high--resolution numerical simulations of
dark matter halos formation (Dubinsky \& Carlberg 1991; Navarro,
Frenk, \& White 1997) seem to point in this direction.  In C98, I
explore further the trend emerged in CP92 and in C96, considering the
analytical DFs of the H0 models and determining the structural and
dynamical limitations imposed to them by dynamical consistency.

\section{The consistency of multi--component systems}

For a multi--component spherical system, where the orbital anisotropy
of each component is modeled according to the OM parameterization, the
DF of the density component $\roi$ is given by:
\begin{equation}
\fid (\qi)={1\over\sqrt{8}\pi^2}{d\over d\qi}\int_0^{\qi} 
{d\roqi\over d\psit}{d\psit\over\sqrt{\qi-\psit}},
\quad
\roqi (r)=\left (1+{r^2\over\rai^2}\right) \roi (r),
\end{equation}
where $\psit (r)=\sum_{\rm k}\psii (r)$ is the total relative potential,
$\qi =\eps-L^2/2\rai^2$, and $0\leq\qi\leq\psit (0)$. $\eps$ and $L$
are respectively the relative energy and the angular momentum modulus
per unit mass, $\ra$ is the {\it anisotropy radius}, and $\fid
(\qi)=0$ for $\qi\leq 0$. If {\it each} $\fid$ is non negative over
all the accessible phase--space, the system is {\it consistent}.
In C92 it was proved that
\smallskip\parn
{\bf Theorem}: A necessary condition (NC) for the non negativity 
of $\fid$ given in eq. (2.1) is:
\begin{equation}
{d\roqi(r)\over dr}\leq 0,\quad 0\leq r \leq\infty .
\end{equation} 
If the NC is satisfied, a {\it strong} (SSC) and a {\it weak
sufficient condition} (WSC) for the non negativity of $\fid$ are
respectively:
\begin{equation}
{d\over dr}\left[{d\roqi(r) \over dr}
{r^2\sqrt {\psit(r)}\over\mt (r)}\right]\geq 0,
\quad
{d\over dr}\left[
{d\roqi(r) \over dr}{r^2\over\mt (r)}\right]\geq 0,
\quad 0\leq r\leq\infty .
\end{equation}
\smallskip\parn
Some considerations follow looking at the previous conditions. The
first is that the violation of the NC is connected only with the
radial behavior of $\roi$ and the value of $\rai$, and so this
condition applies {\it independently} of any other interacting
component added to the model. Even when the NC is satisfied, $\fid$
can be negative, due to the radial behavior of the integrand in
eq. (2.1), which depends on the total potential, on the particular
$\roi$, and on $\rai$; so, a range of permitted values of $\rai$
satisfying the NC must be discarded.  Naturally, the true critical
anisotropy radius is always larger than or equal to that given by the
NC, and smaller than or equal to that given by the SSC (WSC).  To
summarize: a model failing the NC is {\it certainly} inconsistent, and
a model satisfying the SSC (WSC) is {\it certainly} consistent; the
consistency of a model satisfying the NC and failing the SSC (WCS) can
be proved only by direct inspection of the DF.

\section{Results and conclusions}

Both density distributions defining the H0 models belong to the family 
of the $\gamma$ models (Dehnen 1993):
\begin{equation}
\rho (r)={3-\gamma\over 4\pi}
         {M\,\rc\over r^{\gamma}(\rc+r)^{4-\gamma}},\quad 0\leq\gamma <3,
\end{equation}
where $M$ is the total mass and $\rc$ a characteristic scale--length.
The main results obtained in C98 can be summarized as follows:

{\bf (1)} The NC, WSC, and SSC that the model parameters must satisfy,
in order to correspond to an H0 system for which the two physically
distinct components have a positive DF, are analytically derived using
the method introduced in CP92.  Some conditions are obtained for the
wider class of two--component $\gau+\gad$ models (of which the H0
models are a special case).  In particular, it is shown that the DF of
the $\gau$ component in isotropic $\gau+\gad$ models is nowhere
negative, independently of the mass and concentration of the $\gad$
component, whenever $1\leq\gau <3$ and $0\leq\gad\leq\gau$. As an
interesting application of this result, it follows that a BH of any
mass can be consistently added at the center of any isotropic member
of the $\gamma$ models family, when $1\leq\gamma <3$. Two important
consequences follow. The first is that the consistency of isotropic HH
(or H+BH) models proved in C96 using an ``ad hoc'' technique is not
exceptional, but a common property of a large class of two--component
$\gamma$ models: for example, also isotropic two--component Jaffe
[Jaffe 1983, $\gamma=2$ in eq. (3.4)] or Jaffe+BH models can be safely
assembled. The second is that in two--component isotropic models, the
component with the steeper central density distribution is usually the
most robust against inconsistency.

{\bf (2)} It is shown that an analytical estimate of a minimum value
of $\ra/\rc$ for one--component $\gamma$ models with a massive
(dominant) BH at their center can be explicitly found. As expected,
this minimum value decreases for increasing $\gamma$.

{\bf (3)} It is shown that the analytical expression for the DF of H0
models with general OM anisotropy can be found in terms of elliptic
functions; the special cases in which each one of the two density
components are embedded in a dominant halo are also discussed.

{\bf (4)} The region of the parameter space in which H0 models are
consistent is explored using the derived DFs: it is shown that, at
variance with the H component, the $\gamma=0$ component becomes
inconsistent when the halo is sufficiently concentrated, even in the
isotropic case.  This is an explicit example of the negative result
found by CP92 described in the Introduction.

{\bf (5)} The combined effect of halo concentration and orbital
anisotropy is finally investigated. The trend of the minimum value for
the anisotropy radius as a function of the halo concentration is
qualitatively similar in both the components, and to that found for HH
models in C96: a more diffuse halo allows a larger amount of
anisotropy. A qualitatively new behavior is found and explained
investigating the DF of the $\gamma=0$ component in the
halo--dominated case for high halo concentrations. It is analytically
shown the existence of a small region in the parameter space where a
sufficient amount of {\it anisotropy} can compensate the inconsistency
produced by the halo concentration on the structurally analogous --
but isotropic -- case.

{\bf (6)} As a final remark, it can be useful to point out some
general trends that emerge when comparing different one and
two--component models with OM anisotropy, as those investigated
numerically in CP92 and CL97, and analytically in C96 and C98.  The
first common trend is that OM anisotropy produces a negative DF
outside the galaxy center, while the halo concentration affects mainly
the DF at high (relative) energies. The second is that the possibility
to sustain a strong degree of anisotropy is weakened by the presence
of a very concentrated halo.  The third is that in two--component
models, in case of very different density profiles in the central
regions, the component with the flatter density is the most
``delicate'' and can easily be inconsistent: particular attention
should be paid in constructing such models.

\end{document}